\documentclass{kapedbk} 
\def\msun{{\rm M_\odot}}
\def\mh{{M_{\bullet}}}

\def\rh{\rho_{\bullet}}
\def\HH{H$_2$}
\def\gsim{\rlap{\lower 2.5pt
 \hbox{$\sim$}}\raise 1.5pt\hbox{$>$}\;}
\def\lsim{\rlap{\lower 2.5pt
   \hbox{$\sim$}}\raise 1.5pt\hbox{$<$}\;}

\chaptersection 

\upperandlowercase

 \let\footnote\savefootnote

\begin{document}

\articletitle[The First Black Holes]
{The Formation and Evolution of the First Massive Black Holes}

\rhead{The First AGN}

\author{Zolt\'an Haiman}

\affil{Department of Astronomy, 1328 Pupin Laboratories\\ 
Columbia University, New York, NY 10027, USA
\footnote{Partial funding provided by NSF grants AST-0307200 and AST-0307291.}}  
\email{zoltan@astro.columbia.edu}

\author{Eliot Quataert}

\affil{Department of Astronomy, 601 Campbell Hall\\
UC Berkeley, Berkeley, CA 94720, USA
\footnote{Partial funding provided by NASA grant NAG5-12043, NSF grant
AST-0206006, and an Alfred P. Sloan Fellowship.}}
\email{eliot@astron.berkeley.edu}

\begin{abstract}
The first massive astrophysical black holes likely formed at high
redshifts ($z\gsim 10$) at the centers of low mass ($\sim 10^6~\msun$)
dark matter concentrations. These black holes grow by mergers and gas
accretion, evolve into the population of bright quasars observed at
lower redshifts, and eventually leave the supermassive black hole
remnants that are ubiquitous at the centers of galaxies in the nearby
universe.  The astrophysical processes responsible for the formation
of the earliest seed black holes are poorly understood.  The purpose
of this review is threefold: (1) to describe theoretical expectations
for the formation and growth of the earliest black holes within the
general paradigm of hierarchical cold dark matter cosmologies, (2) to
summarize several relevant recent observations that have implications
for the formation of the earliest black holes, and (3) to look into
the future and assess the power of forthcoming observations to probe
the physics of the first active galactic nuclei.
\end{abstract}


\section{Introduction}

It seems established beyond reasonable doubt that some supermassive
black holes (SMBHs) were fully assembled early in the history of the
universe. The handful of bright quasars at $z\gsim 6$ are likely
powered by holes as massive as $\sim10^9~\msun$, and the spectra and
metallicity of these objects appear remarkably similar to their
counterparts at moderate redshifts (Fan et al.\ 2003).  Indeed, if one
selects individual quasars with the same luminosity, their properties
show little evolution with cosmic epoch.\footnote{A possibly important
exception is tentative evidence for increasing Eddington ratios
towards higher redshifts, as discussed in \S~\ref{zoltansubsec:sdss}.}

This implies that the behavior of individual quasars is probably
determined by local physics near the SMBH and is not directly coupled
to the cosmological context in which the SMBH is embedded.  However,
it is clear that the quasar population as a whole does evolve over
cosmic timescales.  Observations from $0\lsim z \lsim 6$ in the
optical (e.g., the Anglo-Australian Telescope's Two Degree Field, or
2dF, and the Sloan Digital Sky Survey, or SDSS) and radio bands
(Shaver et al.\ 1994) show a pronounced peak in the abundance of
bright quasars at $z\approx 2.5$. Recent X-ray observations confirm
the rapid rise from $z=0$ towards $z\approx 2$ for X-ray luminous
sources ($L_X>10^{44}$~ergs~s$^{-1}$; Barger et al.\ 2003) but have
not shown evidence for a decline at still higher redshifts (Miyaji et
al.\ 2000).

The cosmic evolution of quasar black holes between $0 \lsim z \lsim 6$
is likely driven by a mechanism other than local physics near the
hole.  This is reinforced by the fact that the timescale of activity
of individual quasars is significantly shorter than cosmic timescales
at $z\lsim 6$, both on theoretical grounds ($\sim 4\times10^7$ yr, the
e-folding time for the growth of mass in a SMBH, whose accretion
converts mass to radiation with an efficiency of $\epsilon=\dot M
c^2/L_{\rm Edd}\sim 10\%$) and is limited by its own [Eddington]
luminosity), and using the duty cycle of quasar activity inferred from
various observations ($\sim 10^7$ yr, e.g., Martini 2004 and
references therein; see also Haiman, Ciotti \& Ostriker 2004).

In the cosmological context, it is tempting to link the evolution of
quasars with that of dark matter halos condensing in a Cold Dark
Matter (CDM) dominated universe, as the halo population naturally
evolves on cosmic timescales (Efstathiou \& Rees 1988).  Indeed, this
connection has proven enormously fruitful and has resulted in the
following broad picture: the first massive astrophysical black holes
appear at high redshifts ($z\gsim 10$) in the shallow potential wells
of low mass ($\lsim 10^8~\msun$) dark matter concentrations. These
black holes grow by mergers and gas accretion, evolve into the
population of bright quasars observed at lower redshifts, and
eventually leave the SMBH remnants that are ubiquitous at the centers
of galaxies in the nearby universe.

Nevertheless, many uncertainties about this scenario remain. Most
importantly, the astrophysical process(es) responsible for the
formation of the earliest seed black holes (and indeed for the
presence of SMBHs at all redshifts) remain poorly understood.  In this
review, we focus on the emergence of the first generation of black
holes, though many of the important questions are quite general and
apply equally to subsequent generations of black holes.  This review
is organized as follows. In \S~\ref{sec:observations}, we summarize
several relevant recent observations that have implications for early
black holes.  In \S~\ref{sec:first}, we describe theoretical
expectations for the formation and growth of these black holes within
the paradigm of hierarchical CDM cosmologies and also discuss early
black holes in the context of cosmological reionization.  In
\S~\ref{sec:theory}, we ``zoom in'' and consider the local physics of
black hole formation.  In \S~{\ref{sec:future}, we look into the
future, with the goal to assess the power of forthcoming observations
to probe the physics of the first Active Galactic Nuclei (AGN). We
offer our conclusions in \S~\ref{sec:conclude}.

\section{Observational Constraints}
\label{sec:observations}

In this section, we review several recent observations and their
implications for the formation of black holes at high redshifts.

\subsection{High Redshift Quasars in the Sloan Survey}
\label{zoltansubsec:sdss}

The most distant quasars to date have been discovered in the SDSS.
This is perhaps somewhat surprising, since the SDSS is a relatively
shallow survey ($i\sim 22$) capable of detecting only the rarest
bright quasars at redshifts as high as $z\sim 6$.  Nevertheless, the
large solid-angle searched for high redshift quasars to date ($\sim
2800$ square degrees) has yielded a handful of such objects (Fan et
al.\ 2000, 2001, 2003).  The most important properties (for our
purposes) of these sources are that they are probably powered by SMBHs
as large as a few $\times~10^9~\msun$ and they appear to be
indistinguishable from bright quasars at moderate ($z\sim 2-3$)
redshifts, with similar spectra and inferred metallicities. In
addition, a large reservoir of molecular gas is already present, even
in the most distant $(z=6.41)$ source (Walter et al.\ 2003).

In short, these SMBHs and their surroundings appear as ``fully
developed'' as their lower redshift counterparts, despite the young
age ($\lsim 10^9$~years) of the universe at $z\gsim 6$.  These rare
quasars are likely harbored by massive ($\sim 10^{13}~\msun$) dark
matter halos that form out of $4-5\sigma$ peaks of the fluctuating
primordial density field. The large halo mass follows directly from
the space density of these sources (see below and Haiman \& Loeb 2001;
another method to confirm the large halo masses is to study the
expected signatures of cosmological gas infall onto such massive
halos, as proposed by Barkana \& Loeb 2003).  Indeed, the environment
and dynamical history of an individual massive dark matter halo at
$z\sim 6$ and $z\sim 3$ can be similar; it is their abundance that
evolves strongly with cosmic epoch. This is broadly consistent with
the observations: the bright $z\sim 6$ quasars look similar to their
$z\sim 3$ counterparts, but their abundance is much reduced (by a
factor of $\sim 20$).

The fact that these quasars are so rare has important implications.
First, they are likely to be the ``tip of the iceberg'' and
accompanied by much more numerous populations of fainter quasars at
$z\gsim 6$. Pushing the magnitude limits of future surveys should
prove rewarding.  The slope of the luminosity function is expected to
be very steep at $i\sim 22$, and only weak constraints are available
to date from the source counts (Fan et al.\ 2003), and from
gravitational lensing (Comerford, Haiman, \& Schaye 2002; Wyithe \&
Loeb 2002a,b; Richards et al.\ 2004). Combining these two yields the
strongest limit of $-d\log\Phi/d\log L\lsim 3$ (Wyithe 2004).  Second,
the steep slope of the dark halo mass function implies that the masses
of the host halos can be ``measured'' from the abundance quite
accurately (see discussion in \S~\ref{zoltansubsec:fossils}).
Conversely, since small changes in the assumed host halo mass results
in large changes in the predicted abundance, large uncertainties will
remain in other model parameters.  In this sense, fainter, but more
numerous quasars (or lack thereof) can have more constraining power
for models that relate quasars to dark halos (see
\S~\ref{zoltansubsec:deep}).

The most striking feature of the SDSS quasars, however, is the large
black hole mass already present at $z\sim 6$.  This presents
interesting constraints on the growth of these holes and how they are
fueled (see \S~\ref{zoltansubsec:mergers} below).  In the rest of this
section, we critically assess whether the inferred large black hole
masses are robust.

The masses of the black holes powering the SDSS quasars are inferred
by assuming that (1) they shine at the Eddington luminosity with a
bolometric correction identical to that of lower redshift quasars
(this is justified by their similar spectra), and (2) they are neither
beamed nor gravitationally lensed (both of these effects would make
the quasars appear brighter).  These assumptions lead to black hole
masses $\mh\approx(2-5)\times 10^9~{\rm M_\odot}$ for the four $z>6$
quasars known to date.  These are reasonable assumptions, which have
some empirical justification.

The hypothesis that the quasars are strongly beamed can be ruled out
based on their line/continuum ratio. If the quasar's emission was
beamed into a solid angle covering a fraction $f$ of $4\pi$, it would
only excite lines within this cone, reducing the apparent
line/continuum ratio by a factor $f$. However, the SDSS quasars have
strong lines. Haiman \& Cen (2002) found that the line/continuum ratio
of the $z=6.28$ quasar SDSS 1030+0524 is about twice that of the
median value in the SDSS sample at $z>2.25$ (Vanden Berk et
al. 2001). Willott et al.\ (2003) apply this argument to the Mg~II
line of the $z=6.41$ quasar SDSS J1148+5251 and reach a similar
conclusion.

Another important uncertainty regarding the inferred black hole masses
is whether the SDSS quasars may be strongly magnified by gravitational
lensing.  The optical depth to strong lensing along a random line of
sight to $z\sim 6$ is small ($\sim 10^{-3}$; e.g., Kochanek 1998;
Barkana \& Loeb 2000).  Nevertheless, magnification bias can
significantly boost the probability of strong lensing.  If the
intrinsic (unlensed) luminosity function at $z\sim 6$ is steep and/or
extends to faint magnitudes, then the probability of strong lensing
for the SDSS quasars could be of order unity (Comerford, Haiman, \&
Schaye 2002; Wyithe \& Loeb 2002a,b). The overwhelming majority (more
than 90\%) of strong lensing events would be expected to show up as
multiple images with separations at least as large as $0.3''$ (it is
difficult to produce strong magnification without such multiple
images, even in non-standard lensing models; Keeton, Kuhlen, \& Haiman
2004). However, {\em Hubble Space Telescope (HST)} observations of the
highest redshift quasars show no signs of multiple images for any of
the $z~\gsim 6$ sources down to an angle of $0.3''$ (Richards et al.\
2004).  Another argument against strong lensing comes from the large
apparent size of the HII regions around the SDSS quasars (Haiman \&
Cen 2002). For example, the spectrum of the $z=6.28$ quasar SDSS
1030+0524 shows transmitted flux over an $\sim 100$~\AA\ stretch of
wavelength blueward of Ly$\alpha$, corresponding to an $\sim 30$
(comoving) Mpc ionized region around the source.  Provided that this
source is embedded in a neutral intergalactic medium (IGM)---the key
assumption for this constraint (it has some justification, see
below)---it is impossible for an intrinsically faint quasar to produce
such a large HII region, even for a long source lifetime (Haiman \&
Cen 2002).  White et al.\ (2003) derive a similar conclusion for the
quasar J1148+5251, although this source could be magnified by a factor
of approximately a few by lensing (subject to the uncertainty of its
actual redshift, $z=6.37-6.41$).

Finally, whether or not the SDSS quasars are shining at the Eddington
limit is difficult to decide empirically.  Vestergaard (2004)
estimated Eddington ratios in a sample of high redshift quasars using
an observed correlation between the size of the broad line region and
the luminosity of the quasar (the correlation is calibrated using
reverberation mapping of lower redshift objects; e.g. Kaspi et
al. 2000; Vestergaard 2002). She finds values ranging from $\approx
0.1$ to $\gsim 1$, with the $z~\gsim 3.5$ quasars having somewhat
higher $L/L_{Edd}$ than the lower redshift population.  In particular,
Vestergaard estimates $L/L_{Edd} \approx 0.3$ for two of the $z~\gsim
6$ SDSS quasars.  Given the uncertainties in these results, this is
quite consistent with the assumption of near-Eddington accretion.
Note further that in an extended lower redshift $0<z<1$ sample, Woo \&
Urry (2002) also find higher Eddington ratios towards $z=1$, but this
may represent a trend towards higher ratios at higher luminosities.
Whether the trend is primarily with redshift or luminosity is an
important question, but large scatter and selection effects presently
preclude a firm answer.

Inferences about Eddington ratios at high redshifts can also be made
by utilizing models of the quasar population as a whole.  Such models
typically assume the Eddington luminosity at higher redshifts, where
fuel is thought to be readily available (Small \& Blandford 1992;
Haehnelt \& Rees 1993). Several semi-analytic models for the quasar
population (Haiman \& Loeb 1998b; Haehnelt, Natarajan, \& Rees 1998;
Valageas \& Silk 1999; Haiman \& Menou 2000; Kauffmann \& Haehnelt
2000; Wyithe \& Loeb 2003b; Volonteri et al.\ 2003) have found that
Eddington ratios of order unity during most of the growth of the black
hole mass also yield a total remnant SMBH space density at $z=0$ that
is consistent with observations (Magorrian et al.\ 1998; Ferrarese \&
Merritt 2000; Gebhardt et al.\ 2000; Graham et al.\ 2001).  Ciotti \&
Ostriker (2001) have modeled the behavior of an individual quasar and
found that (provided fuel is available) the luminosity is near the
Eddington value during the phases when the quasar is on.  Despite
these arguments, one cannot directly rule out the possibility that the
SDSS quasars shine at super-Eddington luminosities (theoretically,
this is possible in the photon bubble models of Begelman 2002; see
\S~\ref{zoltansubsec:Eddington}). We emphasize that if this were true
and the masses were lower than $10^9~\msun$, then the SDSS quasars
would have to be luminous for only a short time: maintaining the
observed luminosities for $\gsim 10^7$ years with a radiative
efficiency of $\epsilon\equiv L/\dot m c^2=0.1$ would bring the black
hole masses up to values of $10^9~\msun$ anyway.

\subsection{{\em Chandra} and {\em Hubble} Deep Fields}
\label{zoltansubsec:deep}

As discussed above, a relatively shallow but large survey, such as the
SDSS, can discover only the rare AGN at high redshifts.  To constrain
population models, deeper surveys that reveal the ``typical'' sources
are more advantageous.  When completed, the SDSS will have delivered
perhaps $10-20$ $z>6$ quasars, but not more---this is due to the
paucity of quasars as bright as $i\sim 22$ at $z\sim 6$. Furthermore,
the reddest SDSS filter (the $z^\prime$ band) extends only to
$\lambda\approx9500~$\AA, making the survey insensitive to sources
beyond $z\sim 6.5$.  In comparison, deep X-ray observations with {\it
Chandra} and {\it XMM-Newton} (and also near-infrared observations
with {\it HST}) could directly detect SMBHs to redshifts well beyond
the horizon of SDSS, provided that such SMBHs exist.

A SMBH at redshift $z=10$ with mass $\mh=10^8~\msun$ (30 times {\em
lower} than the masses of the $z\sim 6$ SMBHs in the SDSS) would have
an observed flux of $\sim 2\times10^{-16}~{\rm ergs~cm^{-2}~s^{-1}}$
in the soft X-ray band (Haiman \& Loeb 1999b), under the reasonable
assumptions that it shines at the Eddington luminosity and that its
emission has a spectral shape similar to a typical quasar near
redshift $z\sim 2-3$ (with $\sim 3\%$ of the bolometric flux emitted
in the range $0.5 (1+z)$~keV$<E<2 (1+z)$~keV for redshifts $5<z<10$;
e.g., Elvis et al.\ 1994).

Semi-analytic models can be utilized to derive the number and redshift
distribution of quasars at $z>5$ by associating quasar activity with
the dark matter halos that are present at these redshifts (Haiman \&
Loeb 1998b; Kauffmann \& Haehnelt 2000; Wyithe \& Loeb 2003b).  The
predictions of the simplest version of these semi-analytic models
(Haiman \& Loeb 1999b) now appear to be significantly higher (by about
a factor of $\sim 10$) than the number of possible $z>5$ quasar {\it
candidates} (Barger et al.\ 2003; discrepancies had also been noted
earlier by Mushotzky et al.\ 2000, Alexander et al.\ 2001, and
Hasinger 2002).  In the most up-to-date version of such a
semi-analytic model, Wyithe \& Loeb (2003b) find predictions
consistent with (at most) only a few $z>5$ sources in the {\em
Chandra} Deep Field-North (CDF-N).  The main difference in this
updated model (which is what reduces the expected number of high
redshift quasars) is that the assumed quasar lifetime is increased
from $\sim 10^6$ to $\sim 10^7$~years, and the scaling $\mh\propto
M_{\rm halo}$ is modified to $\mh\propto M_{\rm
halo}^{5/3}(1+z)^{5/2}$. These scalings arise due to the radiative
feedback assumed to limit the black hole growth in these models (see
discussion in \S~\ref{zoltansubsec:fossils}).

More generally, the number (or, currently, the upper limit) of high
redshift ($z>6$) sources detected in the CDFs will place the best
constraints to date on quasar evolution models at these high redshifts
(although no systematic assessment of these constraints yet exists in
a suite of models). The current best constraint comes from comparing
model predictions to the dearth of high redshift quasars in the
optical {\it Hubble} Deep Fields (HDFs; Haiman, Madau, \& Loeb 1999).
The results of Haiman \& Loeb (1999) demonstrate that the CDF
constraints are superior to these. Models satisfying the upper limit
from the HDFs (e.g., by postulating that SMBHS do not exist, or are
not fueled, in halos with circular velocities $v~\lsim 75$ km/s) still
result in significant overpredictions for the CDFs.

\subsection{Shedding (Quasar) Light on the Accretion History}
\label{zoltansubsec:soltan}

It has long been proposed that quasar activity is powered by accretion
onto SMBHs (Salpeter 1964; Zel'dovich 1964; Lynden-Bell
1967). It has also been realized that the cumulative radiation output
of all quasars translates into a significant amount of remnant black 
hole mass, presumably to be found at the centers of local galaxies
(Lynden-Bell 1967; So\l tan 1982).  Over the past few years, there has
been renewed interest in this connection because we now have a good
estimate for the total present-day black hole mass density (in addition 
to an estimate of the cumulative quasar light output).  This estimate is
allowed by the recent detection of SMBHs at the centers of several
dozen nearby galaxies (Magorrian et al. 1998), and the tight
correlation of their masses with the masses (Magorrian et al.\ 1998),
velocity dispersions (Ferrarese \& Merritt 2000; Gebhardt et al.\ 2000),
and light profiles (Graham et al.\ 2001) of the spheroids of their host
galaxies.  The connection between quasar light output and remnant mass
provides constraints on the accretion history of the SMBHs and on the
presence of a (yet undetected) population of very high redshift AGN.

Many studies have used this correlation to estimate the total
present-day black hole mass density (e.g., Salucci et al.\ 1999;
Haehnelt \& Kauffmann 2000; Yu \& Tremaine 2002; Haiman, Ciotti, \&
Ostriker 2004). The simplest way to quickly estimate this quantity is
to multiply the local spheroid mass density $\Omega_{\rm
sph}=(0.0018^{+0.00121}_{-0.00085})h^{-1}$ (Fukugita et al.\ 1998) by
the mean ratio $\mh/M_{\rm sph}=10^{-2.9}$ (Merritt \& Ferrarese
2001). The latter ratio is a factor of $\sim4$ times smaller than the
value $10^{-2.28}$ in the original paper (Magorrian et al.\ 1998); the
correction is due mostly to improved models that include velocity
anisotropies (see also van der Marel 1997).  This gives (for $h=0.72$)
$\rh=5\times 10^5~\msun~{\rm Mpc^{-3}}$.  The most sophisticated
analysis to date is by Yu \& Tremaine (2002), who utilize the tight
$\mh-\sigma$ relation and the velocity function of early-type galaxies
measured in SDSS to find $\rh=(3\pm 0.5)\times 10^5~\mh~{\rm
Mpc^{-3}}$.  An additional correlation is observed between black hole
mass and host circular velocity beyond the optical radius (Ferrarese
2002a). This directly ties the black hole to its dark matter halo,
and, although the results depend strongly on the assumed halo profile,
this allows a more refined modeling of the evolution of the quasar
black hole population (e.g., Wyithe \& Loeb 2003b).

The remnant black hole mass density (see Ferrarese, this volume) has
important implications for understanding the first AGN.  Several
authors have compared the cumulative light output of known quasars
(which must be summed over all luminosities and redshifts) with the
remnant black hole mass density.  In general, this yields the average
radiative efficiency of SMBH accretion, with the result $\epsilon\sim
10\%$. Note the paradigm shift: originally, the same analysis was used
to argue that (then unobserved) SMBHs must be ubiquitous in local
galaxies (Lynden--Bell 1969; So\l tan 1982).  Immediately after the
discovery of the local SMBHs, it appeared that there was {\it too
much} local black hole mass, which was taken to imply that most of the
quasar accretion must occur in an optically faint phase (e.g.,
Haehnelt, Natarajan, \& Rees 1998).  The revised, more accurate
estimates of the local SMBH space density (decreased by a factor of
four from the original value) appear consistent with the hypothesis
that the optical quasar population has a mean radiative efficiency of
$\epsilon\sim 10\%$.  It then follows that most of the mass of SMBHs
was accreted during the luminous quasar phase at $z\sim 2-3$, and only
a fraction of the total $\rh$ could have been built during the
formation of the earliest AGN at $z>6$.  However, there are caveats to
this argument: a large (and even dominant) contribution to the total
mass from high redshifts is allowed if the radiative efficiency of the
$z\sim 3$ population is as high as 20\% (and if the high redshift
quasars remain undetectable, either because they are intrinsically
faint or obscured).

The above discussion has focused on optically luminous quasars.  A
significant fraction of black hole growth may, however, occur via
obscured objects, which show up in hard X-ray, infrared, or
submillimeter observations, but not in the optical (see Cowie \&
Barger, this volume). This possibility is strongly suggested by models
of the X-ray background, which require a factor of a few more obscured
AGN than unobscured AGN (see Fabian 2004 for a review).  The precise
fraction of black hole growth that occurs in an optically obscured
phase is uncertain by a factor of a few. For example, the local black
hole mass density has about a factor of two uncertainty, depending on
the details of which $M-\sigma$ correlation is used and how it is
extrapolated to the entire galaxy sample in the universe (Yu \&
Tremaine 2002; Ferrarese 2002b). This immediately allows for a
comparable amount of obscured and unobscured accretion with a typical
efficiency of 10\%.  If, however, most black holes are rapidly
rotating, or if magnetic torques are important at the last stable
orbit (e.g., Gammie 1999), then the mean efficiency could be
significantly larger ($\sim 40$\%).  In this case, most accretion may
occur in an optically obscured phase.  Hard X-ray (e.g., the NuSTAR
telescope recently selected by NASA for Phase A study as a SMEX
mission) and infrared (e.g., {\it Spitzer Space Telescope})
observations are required to provide an unbiased view of the growth of
SMBHs.

\subsection{Local Black Holes as Fossils}
\label{zoltansubsec:fossils}

As mentioned above, SMBHs appear ubiquitous in local galaxies, with
their masses correlating with the global properties of their host
spheroids. Several groups have noted the broad natural implication
that the formation of the SMBHs and their host spheroids must be
tightly linked (see, e.g., Monaco et al.\ 2000; Kauffmann \& Haehnelt
2001; Granato et al.\ 2001; Ciotti \& van Albada 2001; Cattaneo et
al.\ 2003; Haiman, Ciotti, \& Ostriker 2004).  Various independent
lines of evidence suggest that spheroids are assembled at high
redshifts ($z\sim 2$; see Cattaneo \& Bernardi 2003 for the recent age
determinations from the Sloan sample and references to older work),
which would be consistent with most of the SMBH mass being accreted
around this redshift (coinciding with the peak of the activity of
luminous quasars, as discussed in \S~\ref{zoltansubsec:soltan}).  This
then has the unwelcome (but unsurprising) implication that the local
SMBHs may contain little direct evidence of the formation of their
seeds at $z>6$.  Indeed, it seems most plausible that the observed
tight correlations, such as between $\mh$ and $\sigma$, are
established by a feedback process which operates when most of the
black hole mass is assembled.  However, the significance of this
hypothesis is that---with the identification of a specific feedback
mechanism---physically motivated extrapolations can be made towards
high redshifts.

Another interesting {\em observational} question is whether the local
$\mh-\sigma$ relation holds at higher redshifts, both in normalization
and in slope (as discussed by several authors), and also in range
(which has received less attention, but see Netzer 2003 and discussion
below). The highest redshift SDSS quasars do appear to approximately
satisfy the $\mh-\sigma$ relation of the local SMBHs.  If $\mh$ is
estimated assuming the Eddington luminosity, and $\sigma$ is estimated
from the circular velocity of the host dark matter halos with the
right space density (e.g., Haiman \& Loeb 2001), then the SDSS quasars
are within the scatter of the $\mh-\sigma$ relations of Gebhardt et
al.\ (2000) and also of Ferrarese (2002a). As explained in
\S~\ref{zoltansubsec:sdss}, the mass inference is reasonable.  The
determination of the halo mass and circular velocity from the observed
abundance of quasars is also more robust than it may at first
appear. This is because, despite the dependence on the poorly known
duty cycle, the halo mass function is exponentially steep for the
massive $M\sim 10^{13}~\msun$ halos at $z\sim 6$; therefore, the
dependence of the inferred halo mass on the duty cycle (and other
uncertainties in the estimated halo abundance) is only logarithmic.
The weakest link in the argument is associating the spheroid velocity
dispersion with the circular velocity of the dark matter halo.
Ferrarese (2002a) shows evidence of a correlation between $\mh$ and
$\sigma$, with the velocity dispersion measured in the dark matter
dominated region of SMBH host galaxies; this establishes a direct link
to the dark halo and puts the above argument on somewhat firmer ground
(although there are still large errors in the inferred correlation,
depending on the halo profile one adopts to convert the measured
circular velocity to total halo mass).

The (tentative) evidence that high redshift AGN also satisfy the
$\mh-\sigma$ relation further supports the idea that the
formation of SMBHs and their host galaxies must be tightly coupled by
cosmology-independent physical processes (since the SDSS quasars are
the rare peaks that have already formed at $z\sim 6$ instead of at 
$z\sim 2$).  Netzer (2003) raises the point that besides the slope and
normalization of the $\mh-\sigma$ relation, the {\it range} (of
masses and velocity dispersions) over which observed galaxies satisfy
this relation has to match between low and high redshifts. In
particular, the largest $\gsim 10^{10}~\msun$ black holes observed 
at high redshifts should also exist at low redshifts, but have not 
yet been discovered.

There have been several suggestions in the literature for the nature
of the dynamical coupling between the formation of the black hole and
its spheroid host.  The most promising is probably radiative or
mechanical feedback from the SMBH on the gas supply in the bulge (Silk
\& Rees 1998; Haehnelt, Natarajan, \& Rees 1998; Blandford 1999; King
2003; Wyithe \& Loeb 2003b).  The essential idea is that when the
black hole in the center of the galaxy grows too large, its outflows
and radiation unbind the gas in the bulge or in the disk, quenching
further black hole growth via accretion and further star formation.
Competition with star formation for the gas supply may also play a
role (Di Matteo et al.\ 2003).  Note that these mechanisms can readily
work at any redshift.

Alternative possibilities for the origin of the $\mh-\sigma$ relation
include: (1) Filling the dark matter loss cone (Ostriker 2000).  In
this model, the growth of the SMBH occurs first through the accretion
of collisional dark matter particles, and subsequently through the
scattering of these particles into orbits that are then perturbed to
pass sufficiently close to the black hole's Schwarzschild radius to be
captured. This model runs into difficulties with the So\l tan argument
discussed in \S~\ref{zoltansubsec:deep}; since the SMBHs are fed
mostly dark matter rather than gas, there is no associated
radiation. (2) Direct capture of stars on high eccentricity orbits by
the SMBH (Zhao, Haehnelt, \& Rees 2002, Merritt \& Poon 2004). This
model has a similar problem because black holes more massive than
$\gsim 10^8~\msun$ do not tidally disrupt stars, so there is again no
radiative output associated with the black hole growth. (3) Stellar
captures by the accretion disk feeding the hole (Miralda-Escud\'e \&
Kollmeier 2004).

\section{First Structure Formation}
\label{sec:first}

In this section, we sketch some basic theoretical arguments relevant
to the formation of structure in the universe. We then discuss
formation mechanisms for SMBHs.

\subsection{Cosmological Perturbations as the Sites of the First 
Black Holes}
\label{zoltansubsec:cosmo}

Recent measurements of the Cosmic Microwave Background (CMB)
temperature anisotropies by the {\em Wilkinson Microwave Anisotropy
Probe (WMAP)}, determinations of the luminosity distance to distant
type Ia Supernovae, and other observations have led to the emergence
of a robust ``best fit'' cosmological model with energy densities in
CDM and ``dark energy'' of $(\Omega_{\rm M},\Omega_{\rm
\Lambda})\approx (0.3,0.7)$ (e.g., Spergel et al. 2003).

The growth of density fluctuations and their evolution into nonlinear
dark matter structures can be followed in this cosmological model from
first principles by semi-analytic methods (Press \& Schechter 1974;
Sheth et al.\ 2001).  More recently, it has become possible to derive
accurate dark matter halo mass functions directly in large
cosmological N-body simulations (Jenkins et al.\ 2001).  Structure
formation in a CDM dominated universe is ``bottom-up'', with low mass
halos condensing first.  Dark matter halos with the masses of globular
clusters ($10^{5-6}~\msun$) are predicted to have condensed from $\sim
3\sigma$ peaks of the initial primordial density field as early as
$\sim1\%$ of the current age of the universe, or at redshifts of
$z\sim 25$.

It is natural to identify these condensations as the sites where the
first astrophysical objects, including the first AGN, were born.  The
nature of the objects that form in these early dark matter halos is
currently one of the most rapidly evolving research topics in
cosmology.

\subsection{Chemistry and Gas Cooling at High Redshifts}
\label{zoltansubsec:H2}

Baryonic gas that falls into the earliest nonlinear dark matter halos
is shock heated to the characteristic virial temperatures of a few
hundred Kelvin. It has long been pointed out (Rees \& Ostriker 1977;
White \& Rees 1978) that such gas needs to lose its thermal energy
efficiently (within about a dynamical time) in order to continue
contracting, or in order to fragment. In the absence of any
dissipation, it would simply reach hydrostatic equilibrium and would
eventually be incorporated into a more massive halo further down the
halo merger hierarchy.  While the formation of nonlinear dark matter
halos can be followed from first principles, the cooling and
contraction of the baryons, and the ultimate formation of stars or
black holes in these halos, is much more difficult to model ab initio.

The gas content of a cosmological perturbation can contract together
with the dark matter only in dark halos above the cosmological Jeans
mass, $M_{\rm J}\approx 10^4~\msun [(1+z)/11]^{3/2}$, in which the
gravity of dark matter can overwhelm thermal gas pressure.  In these
early, chemically pristine clouds, radiative cooling is dominated by
${\rm H_2}$ molecules. As a result, gas phase ${\rm H_2}$
``astrochemistry'' is likely to determine the epoch when the first AGN
appear (the role of ${\rm H_2}$ molecules for early structure
formation was reviewed by Abel \& Haiman 2001).  Several papers have
constructed complete gas-phase reaction networks and identified the
two possible ways of gas-phase formation of ${\rm H_2}$ via the ${\rm
H^-}$ or ${\rm H_2^+}$ channels.  These were applied to derive the
${\rm H_2}$ abundance under densities and temperatures expected in
collapsing high redshift objects (Hirasawa 1969; Matsuda et al.\ 1969;
Palla et al.\ 1983; Lepp \& Shull 1984; Shapiro \& Kang 1987; Kang et
al.\ 1990; Kang \& Shapiro 1992; Shapiro, Giroux, \& Babul 1994).
Studies that incorporate ${\rm H_2}$ chemistry into cosmological
models and that address issues such as non-equilibrium chemistry,
dynamics, or radiative transfer have appeared relatively more
recently. Haiman, Thoul, \& Loeb (1996) used spherically symmetric
simulations to study the masses and redshifts of the earliest objects
that can collapse and cool via ${\rm H_2}$; their findings were
confirmed by a semi-analytic treatment in Tegmark et al.\ (1997).  The
first three dimensional cosmological simulations that incorporate
${\rm H_2}$ cooling date back to Gnedin \& Ostriker (1996, 1997) and
Abel et al.\ (1997).

The basic picture that emerged from these papers is as follows.  The
${\rm H_2}$ fraction after recombination in the smooth
``protogalactic'' gas is small ($x_{\rm H2}=n_{\rm H2}/n_{\rm H}\sim
10^{-6}$). At high redshifts ($z~\gsim 100$), ${\rm H_2}$ formation is
inhibited, even in overdense regions, because the required
intermediaries ${\rm H_2^+}$ and H$^-$ are dissociated by cosmic
``microwave''background (CMB, but with the typical wavelength in the
infrared) photons.  However, at lower redshifts, when the CMB photons
redshift to lower energies, the intermediaries survive, and a
sufficiently large ${\rm H_2}$ abundance builds up inside collapsed
clouds ($x_{\rm H2}\sim 10^{-3}$) at redshifts $z~\lsim 100$ to cause
cooling on a timescale shorter than the dynamical time.  Sufficient
\HH\ formation and cooling is possible only if the gas reaches
temperatures in excess of $\sim 200$~K or masses of a few
$\times~10^5~\msun [(1+z)/11]^{-3/2}$ (note that while the
cosmological Jeans mass increases with redshift, the mass
corresponding to the cooling threshold, which is well approximated by
a fixed virial temperature, has the opposite behavior and decreases at
high redshift).  The efficient gas cooling in these halos suggests
that the first nonlinear objects in the universe were born inside
$\sim 10^5~\msun$ dark matter halos at redshifts of $z\sim 20$
(corresponding to an $\sim3\sigma$ peak of the primordial density
peak).

The behavior of metal-free gas in such a cosmological ``minihalo'' is
a well posed problem that has recently been addressed in three
dimensional numerical simulations (Abel, Bryan, \& Norman 2000, 2002;
Bromm, Coppi, \& Larson 1999, 2002).  These works have been able to
follow the contraction of gas to much higher densities than previous
studies. They have shown convergence towards a temperature/density
regime of ${\rm T \sim 200~{\rm K}}$, ${\rm n \sim 10^{4}~{\rm
cm}^{-3}}$, dictated by the critical density at which the excited
states of ${\rm H_2}$ reach equilibrium and cooling becomes less
efficient (Galli \& Palla 1998).  The 3D simulations suggest that the
mass of the gas does not fragment further into clumps below sizes of
$10^{2}-10^{3}~\msun$, but rather it forms unusually massive stars.
Such stars would naturally leave behind black hole seeds, which can
subsequently grow by mergers and accretion into the SMBHs.
Interestingly, massive stars have an ``either/or''
behavior. Nonrotating stars with masses between $\sim 40-140~\msun$
and above $\sim 260~\msun$ collapse directly into a black hole without
an explosion, and hence without ejecting their metal yields into the
surrounding medium, whereas stars in the range $\sim 140-260~\msun$
explode without leaving a remnant (Heger et al.\ 2003). This dichotomy
is especially interesting because early massive stars are attractive
candidates for polluting the IGM with metals at high redshifts (Madau,
Ferrara, \& Rees 2001; Wasserburg \& Qian 2000).  It is likely that
the first stars had a range of masses, in which case they could
contribute to both metal enrichment and to the seed black hole
population, with a relative fraction that depends sensitively on their
initial mass function (IMF).

\subsection{Cosmological Reionization: Do the First Black Holes Contribute?}
\label{zoltansubsec:reionization}

Perhaps the most conspicuous effect of the first generation of light
sources, once they collectively reach a critical emissivity of
ionizing radiation, is the reionization of the IGM.  As has long been
known, the absence of strong HI Ly$\alpha$ absorption (i.e., a
so-called Gunn-Peterson trough, Gunn \& Peterson 1965, hereafter GP)
in the spectra of distant sources implies that the IGM is highly
ionized (with volume averaged neutral fractions $\lsim 10^{-4}$) at
all redshifts $z\lsim 6$.  There have been two observational
breakthroughs recently. On the one hand, there is evidence, from the
strong absorption in the spectra of the highest redshift SDSS quasars,
that the transition from a neutral to a highly ionized state of the
IGM is occurring close to $z\sim 6$ (Becker et al.\ 2001; Fan et al.\
2003; White et al.\ 2003).\footnote{The above statement refers to a
transition from neutral to ionized hydrogen.  A transition from HeII
to HeIII appears to be occurring at a lower redshift, $z\sim 3$ (e.g.,
Heap et al.\ 2000).}  On the other hand, the recent detection of a
large electron scattering optical depth by the {\it WMAP} satellite
implies that significant ionization had taken place at much higher
redshifts ($z\sim 15$, Spergel et al.\ 2003).  There is currently a
flurry of activity trying to interpret these results in the context of
reionization models (see Haiman 2004 for a recent review). The
electron scattering optical depth measured by {\it WMAP} still has a
significant uncertainty, $\tau=0.17\pm 0.04$ (Kogut et al.\ 2003;
Spergel et al.\ 2003).  Nevertheless, these developments bring into
sharp focus an interesting ``old'' question: could AGN have
contributed to the reionization of the IGM?  A natural follow-up
question would then be, can we use reionization as a probe of the
earliest AGN?  In this section, we highlight the need for an early
population of ionizing sources and assess whether these could be early
AGN. We start with a critical review of the recent observations and
their implications.

$\bullet$ {\em Reionization and the Gunn--Peterson Troughs.} 
At the time of this writing, there are four known quasars at $z>6$
(Fan et al.\ 2003).  This redshift appears to coincide with the tail
end of the epoch of reionization.  In the spectra of about a dozen
bright quasars at $z>5$, the amount of neutral hydrogen absorption
increases significantly (by about an order of magnitude) from
$z\sim5.5$ to $z\sim6$ (McDonald \& Miralda-Escud\'e 2001; Cen \&
McDonald 2002; Fan et al.\ 2002; Lidz et al.\ 2002), with the highest
redshift quasars showing full Gunn-Peterson troughs consistent with
no transmitted flux.\footnote{White et al.\ (2003) detect flux blueward
of Ly$\alpha$ in the spectra of the $z=6.41$ quasar, but they
attribute this flux to an intervening $z=4.9$ galaxy. 
Even if the feeble flux originated from the quasar, the strong lower
limits on the mean neutral fraction would remain.}  Assuming
photoionization equilibrium, this corresponds to a sharp increase,
by at least an order of magnitude, in the H-ionizing background
radiation from $z\sim 6$ to $z\sim 5.5$.

The actual numerical limit on the mean mass (volume) weighted neutral
fraction at $z\sim 6$ is only $x_{\rm HI}>10^{-2}$ ($x_{\rm
HI}>10^{-3}$), which does not directly establish that we are probing
the neutral epoch of the IGM with $x_{\rm HI}\sim 1$.  However, the
observed steep evolution of the ionizing background suggests that this
is the case.  Note that the ionizing background scales as $J\propto
\epsilon \lambda$, where $\epsilon$ is the emissivity (per unit
volume), and $\lambda$ is the mean free path of ionizing photons.
Specifically, the increase from $z\sim 6$ to $z\sim 5.5$ is much
steeper than the evolution of the emissivity of the known galaxy
population (e.g., Madau \& Pozzetti 2000).  The emissivity of the
bright quasar population evolves faster, but they cannot account for
the background needed to cause the observed high level of ionization
(Shapiro, Giroux \& Babul 1994; Haiman, Abel \& Madau
2001). Alternatively, the observed steep evolution of $J$ could come
from a rapid evolution of the mean free path.  At lower redshifts, the
mean free path is dominated by the poorly known abundance of Lyman
limit systems (Haardt \& Madau 1996) and evolves
rapidly. Nevertheless, this evolution, scaling approximately as
$\lambda_{\rm mfp}\propto (1+z)^{-6}$ (Cen \& McDonald 2002), still
accounts only for a factor of about two change in $J$ from $z\sim 5.5$
to $z\sim 6$.

It therefore appears that the steep evolution of $J$ seen in the SDSS
quasar spectra is difficult to understand without invoking some
additional physical effect(s).  On the other hand, the steep evolution
would be naturally expected if we were detecting the end of the
reionization epoch. Before the discrete ionized bubbles around
isolated sources first overlap, each hydrogen atom is exposed to at
most a single (or a few, if the sources are clustered) ionizing
sources.  However, after the bubbles percolate, the mean free path to
ionizing photons undergoes a sudden drop, and the background seen by a
typical atom will sharply increase to a sum over the entire ``Olbers'
integral'', dominated by numerous distant sources (e.g., Haiman \&
Loeb 1999; Gnedin 2000).  The timescale for the build-up of the
background, the light crossing time across individual HII bubbles, is
a small fraction of the Hubble time. This would be the most economical
explanation of the observed steep evolution of $J$.

Note that Songaila \& Cowie (2002) and Songaila (2004) reach a
different conclusion, and find that the spectra are consistent with a
smoothly evolving ionizing background over this redshift interval.
The strongest conclusions about whether or not there is a sudden
change in the ionizing background at $z\sim 6$ can be drawn from the
level of Ly$\gamma$ and Ly$\beta$ absorption. Because of the smaller
oscillator strengths of these higher transitions, the IGM is less
opaque in these lines than in the corresponding Ly$\alpha$
line. Hence, these higher lines yield a stronger lower limit on the
neutral fraction and a stronger upper limit on the ionizing
background.  Quantifying the value of the corresponding Ly$\alpha$
opacity and of the implied ionization rate, from the observed
transmission in the Ly$\beta$ and Ly$\gamma$ region, is complicated,
and must involve a detailed modeling of the underlying density
distribution. It appears that the differences between Songaila \&
Cowie's conclusions, and that of the other groups are in the treatment
of the higher lines (Songaila 2004). Given the significance of whether
or not we are detecting the reionization epoch, this will be important
to clarify in future work.

$\bullet$ {\em Reionization and the CMB.}  
The free electrons produced by reionization scatter a fraction of the
background CMB photons, offering an alternative way of probing the
reionization history. (This approach probes to redshifts much
higher than can currently be studied spectroscopically.)  The
scattering of the photons damps the temperature fluctuations on small
angular scales (i.e., on scales below the apparent size of the horizon
at reionization, corresponding to spherical index $\ell~\gsim 10$;
this is a purely geometrical effect) but boosts the ``primordial''
polarization signal at large angles ($\ell~\lsim 10$; see Zaldarriaga
1997).  The effect on the temperature anisotropies is essentially an
overall suppression of power on the scales where it is measurable, and
it is therefore unfortunately nearly degenerate with the intrinsic
amplitude of the fluctuation power spectrum. However, Kaplinghat et
al.\ (2003) showed that one can break this degeneracy, and detect the
reionization signature, in the polarization power spectrum.  It is
indeed in the ``TE'' (temperature--polarization angular
cross--correlation) map that the {\it WMAP} team discovered evidence of
early reionization and measured the value $\tau=0.17\pm 0.04$ for the
electron scattering optical depth (Spergel et al.\ 2003; Kogut et
al.\ 2003).  Assuming a single step reionization (sudden transition
from a neutral to a fully ionized IGM at redshift $z_r$), this
translates to a reionization redshift of $z_r=17\pm 4$.

$\bullet$ {\em Reionization and Early Black Holes.}
The above {\it WMAP} result is inconsistent at the $3\sigma$ level
with a sudden percolation of HII bubbles occurring at $z\sim 6$, which
would correspond to the low optical depth of $\tau=0.04$.  This
discrepancy is reduced (to the $\sim 2\sigma$ level) even in the
simplest models for reionization in which the ionizing emissivity
traces the collapse of DM structures.  With a reasonable choice of
efficiency parameters in such a model, percolation indeed occurs
around $z\sim 6$, satisfying the GP trough detections. In such models,
there is a natural ``tail'' of partial ionization, extending to
redshifts beyond the percolation epoch, which predicts the total
$\tau\sim 0.08$ (Haiman \& Holder 2003; Ciardi et al. 2003).  However,
if the high value of $\tau=0.17$ is confirmed in future CMB
polarization data (e.g., by several additional years of {\it WMAP}
data), the implication will remain: there are additional sources of
ionizing radiation at $z\sim 15$.  Most importantly, with further
improved CMB polarization measurements by {\it Planck}, the
reionization history at high redshifts can be mapped to high
precision (Kaplinghat et al.\ 2003; Holder et al.\ 2003).

The emissivity of the bright optical quasar population drops steeply
at high redshifts ($z~\gsim 3$; e.g., Fan et al.\ 2002).  There is a hint
that the evolution towards high redshifts is flatter in X-rays (Miyaji
et al.\ 2000). While this could be explained if optical quasars were
selectively more dust-obscured at high redshifts, this interpretation
would fail to explain the sharp decline towards high redshifts that is 
also seen in the radio (Shaver et al.\ 1996).  If the sharp decline is 
real, it is easy to show that quasars do not contribute significantly 
to the ionizing background at $z~\gsim 6$ (Madau, Haardt, \& Rees 1999; 
Haiman, Abel, \& Madau 2001; Fan et al.\ 2001; Barger et al.\ 2003) 
and thus cannot account for the GP troughs detected in the SDSS quasars 
at this redshift.

However, there is, at least in principle, still room for AGN to
contribute to reionization.  First, the above ignores the possible
presence of faint ``miniquasars'' (a terminology introduced by Haiman,
Madau, \& Loeb 1999) below the current detection thresholds. It has
been pointed out (Haiman \& Loeb 1998b; Haehnelt, Natarajan, \& Rees
1998) that there could be a significant population of such faint
quasars and that their expected abundance depends crucially on the
duty cycle of quasar activity.  If the quasar lifetime is short
($\lsim 10^7$~years), then quasars must reside in intrinsically
abundant, low mass halos in order to match their observed surface
density on the sky. Conversely, if quasars are long-lived, they must
be harbored by the rarer, more massive halos (for the same apparent
abundance).  The abundance of low mass halos declines less rapidly
(and can even increase for $M_{\rm halo}< M_*$) towards high
redshifts, and therefore if the quasar duty cycle is short, one
expects a larger number of yet-to-be detected ``miniquasars''.
Quasar lifetimes are currently uncertain but are constrained to lie
in the range $10^6-10^8$~years (see the review by Martini 2004). A
particularly relevant method to obtain the lifetime (and thus host
halo mass) for the typical quasar at a fixed luminosity is to study
the spatial clustering of quasars in large surveys such as 2dF or SDSS
(Haiman \& Hui 2001; Martini \& Weinberg 2001). Current results from
2dF favor $t\lsim 10^7$~years (see Croom et al.\ 2004).

In the simple models of Haiman \& Loeb (1998) that assume a short
quasar lifetime, quasars can reionize the IGM by $z~\gsim 10$. That
model was ``calibrated'' to reproduce the original observed relation
between SMBH mass and bulge mass at $z\sim 0$ by Magorrian et
al.\ (1998).  However, the model runs into difficulties with more
recent observations: (1) it overproduces the expected counts of faint
X-ray sources in the CDFs, and (2) it is no
longer consistent with the more recent local SMBH mass estimates
(which are reduced by a factor of $\sim$4) and their steeper
dependence on the velocity dispersion $\mh\propto\sigma^{4-5}$
rather than $\propto \sigma^3$.  Wyithe \& Loeb (2003b) recently
presented an updated model satisfying these constraints.  In their
model, the abundance of faint quasars at high redshifts falls short
of reionizing the universe at $z\sim 6$.

Despite the above conclusions, it is natural to ask whether the
abundance of fainter miniquasars could be higher, and whether they
could then significantly contribute to the reionization history.
Ricotti \& Ostriker (2004) show that such SMBHs can significantly
ionize the universe if they contain a fraction $\gsim 10^{-5}$ of all
baryons. Another example is a large population of intermediate ($\sim
100~\msun$) black holes, which have a harder spectrum and are more
efficient ionizers (Madau et al.\ 2004).  A general constraint on the
quasar contribution to reionization comes from the delay between HI
and HeII reionization epochs.  Quasars have a hard spectrum, with the
ratio of the number of photons above 4 and 1 Rydbergs about $\sim
10\%$, roughly the ratio of He versus H atoms (the spectra of
intermediate mass black holes (IMBHs) are harder, and they produce a
factor of $\sim$ two more He-ionizing photons). One may naively expect
that hydrogen and helium would be reionized simultaneously, in
contrast with the observed redshift $z_r{\rm (H)}~\gsim 6$ and $z_r
{\rm (HeII)}\sim 3$.  Including the fact that HeII recombines $\sim 5$
times faster than HI would translate to a delay that is consistent
with the values above (Miralda--Escud\'e \& Rees 1993).  However, a
delay from $z_r\rm{(H)}\sim 15$ to $z_r\rm{(HeII)}\sim 3$ would be
inconsistent with a pure miniquasar reionization scenario.  This means
that if hydrogen reionization was caused by miniquasars at $z\sim 15$,
then HeII was likely reionized around the same redshift; it then
subsequently recombined (as stellar sources overtook miniquasars as
the dominant ionizing sources) and was reionized again by the bright
quasar population at $z\sim 3$.  This non-trivial evolution implies a
complex thermal history of the IGM, which may leave detectable
imprints on its temperature distribution at lower redshifts (Hui \&
Haiman 2003).

The hard spectra of quasars produce several other distinguishing
features for reionization (Oh 2001; Venkatesan et al.\ 2001).  Because
the mean free path is longer than the Hubble length for photons with
energies $\gsim [(1+z)/10]^{1/2}$~keV, there is no sharp ``edge'' for
the discrete HII regions surrounding the ionizing sources. As a
result, the neutral fraction should decrease gradually throughout most
of the entire IGM. This is in sharp contrast with the Swiss cheese
topology of reionization by softer photons.  Furthermore, X-ray
photons deposit a significant fraction ($\sim 1/3$) of their energy
into ionizations only, while the IGM is close to neutral. Once the
ionized fraction reaches $\sim 30\%$, most of their energy is
thermalized with the electrons (e.g., Shull \& van Steenberg 1985).  As
a result, reionization by quasars would be quite different from the
stellar case: the IGM would be gradually ionized to the ionized
fraction of $\sim 30\%$ (as opposed to suddenly fully ionized).  These
features make it unlikely that quasars contributed significantly to the
sudden elimination of the GP troughs at $z\sim 6$.  However, the same
features would be attractive in producing partial reionization at high
redshifts, and thus would help in explaining the large optical depth 
measured by {\it WMAP} (Madau et al.\ 2004; Ostriker et al.\ 2003).  
Note that in
this scenario, normal stars would ``take over'' and dominate the
ionizing background at $z\sim 6$, causing the overlap of highly
ionized regions.  The stars would then concurrently heat the IGM to
$\sim 2\times10^4$~K.  Hui \& Haiman (2003) have argued (see also
Theuns et al.\ 2002) that the IGM could not be kept {\it fully} ionized
continuously from $z=15$ to $z=4$ because it would then cool
adiabatically to a temperature that is below the observed value at
$z\sim 4$.  The above scenario could naturally avoid this constraint.

We have therefore seen that the first AGN at $z>6$ could, in
principle, still be important contributors to reionization at high
redshifts.  To conclude this section, we point out yet another
potential constraint.  At energies above $\sim 1$~keV (rest frame at
$z=0$), there is little absorption, and whatever radiation was
produced by the high redshift quasar population would add cumulatively
to the present-day background.  Most of the soft X-ray background has
already been resolved into low redshift sources (Mushotzky et al.\
2000; see also Wu \& Xue 2001 and references therein).  Dijkstra,
Haiman, \& Loeb (2004a) find that the putative high redshift quasars,
if they are to fully reionize the IGM, would overproduce the soft
X-ray background.  However, distant miniquasars that produce enough
X--rays to only partially ionize the IGM to a level of at most
$x_e\sim 50\%$ are still allowed.

\section{Massive Black Hole Formation}
\label{sec:theory}

Having reviewed the general problem of structure formation at high
redshifts, we now focus on the poorly understood question of how SMBHs
were assembled in the first place. This is an outstanding problem, and
it is not even clear whether the first nonlinear objects in the
universe were stars or black holes, and whether galaxies or their central 
black holes formed first (see below).  
The leading ideas related to the formation
of SMBHs at high redshifts can be broadly divided into three areas: (1)
formation of seed black holes from ``normal'' stellar evolution and
subsequent accretion to form SMBHs, (2) direct collapse of gas to a
SMBH, usually via a supermassive star/disk, and (3) formation of a SMBH 
(or an IMBH seed) by stellar dynamical processes
in dense stellar systems, such as star clusters or galactic nuclei.  
It is, of course, likely that all of these processes could be relevant
(e.g., Begelman \& Rees 1978; Rees 1984).

\subsection{Seed BHs and Accretion}
\label{zoltansubsec:Eddington}

In view of the evidence described in \S~\ref{zoltansubsec:sdss}, it is
quite convincing that the SDSS quasars at $z\sim 6$ have masses of
several $10^9~\msun$.  Having black holes as massive as this at such
an early stage in the evolution of the universe requires explanation.
The simplest possibility is that they grow by gas accretion out of a
stellar mass seed black hole, left behind by an early massive star.
The earliest stars, forming out of metal free gas, are thought to be
massive (several $100~\msun$; Abel, Bryan, \& Norman 2000, 2002;
Bromm, Coppi, \& Larson 1999, 2002).  Such stars can leave behind a
substantial fraction of their original mass as a black hole (Heger et
al.\ 2003; Carr, Bond, \& Arnett 1984).  As emphasized by Haiman \&
Loeb (2001), if the subsequent gas accretion obeys the Eddington limit
and the quasar shines with a radiative efficiency of 10\%, then the
time it takes for a SMBH to grow to the size of $3\times 10^9~\msun$
from a stellar seed of $\sim 100~\msun$ is $3\times 10^7~{\rm
ln}(3\times10^9/100)~{\rm yr}\sim 7\times 10^8~{\rm yr}$.  This is
comparable to the age of the universe at $z=6$ ($\sim 9\times
10^8~{\rm yr}$ for a flat $\Lambda$CDM universe with $H_0=70~{\rm
km~s^{-1}~Mpc^{-1}}$ and $\Omega_M=0.3$).  Therefore, the presence of
these black holes is consistent with the simplest model for black hole
growth, {\em provided that the seeds are present early on}, at $z\gsim
15$.

In this context, the crucial question is whether gas can accrete at a
highly super-Eddington rate onto a black hole, i.e., with $\dot M \gg
\dot M_{Edd}$, where $\dot M_{Edd} = 10 L_{Edd}/c^2 \approx 1.7
M_8~\msun$~yr$^{-1}$ is the accretion rate that would produce an
Eddington luminosity if accretion onto a black hole of mass $10^8
M_8~\msun$ proceeded with $10\%$ radiative efficiency. If so, this
could lead to rapid black hole growth at high redshifts.  Constraints
on BH seeds and their formation redshifts would therefore be much less
stringent.  If mass is supplied to a black hole at $\dot m \equiv \dot
M/\dot M_{Edd} \gg 1$, the photons are trapped in the inflowing gas
because the photon diffusion time out of the flow becomes longer than
the time it takes the gas to accrete into the black hole (e.g.,
Begelman 1978; Begelman \& Meier 1982).  The resulting accretion is
thus not via the usual thin disk (Shakura \& Sunyaev 1973), but rather
via a radiatively inefficient flow (RIAF); the luminosity is still set
by the Eddington limit, but most of the gravitational binding energy
released by the accretion process is not radiated away (being trapped
in the flow).

The growth of SMBHs at high redshifts probably proceeds via an
optically thick photon trapped accretion flow with $\dot m \gg 1$.
Indeed, it would be a remarkable coincidence if the mass supply rate
were precisely $\sim \dot M_{Edd}$ (required for a thin accretion
disk) during the entire growth of massive black holes.  It is more
likely that the mass supply rate is initially much larger in the dense
environments of high redshift galaxies ($\dot m \gg 1$) and then
slowly decreases with time as the galaxy is assembled (e.g., Small \&
Blandford 1992; Cavaliere, Giacconi, \& Menci 2000).  Recent
theoretical calculations imply that if $\dot m \gg 1$, very little of
the mass supplied to the black hole actually reaches the horizon; most
of it is driven away in an outflow (see, e.g., simulations of RIAFs by
Stone, Pringle, \& Begelman 1999; Igumenshchev \& Abramowicz 1999;
Stone \& Pringle 2001; Hawley \& Balbus 2002; Igumenshchev et
al. 2003; Proga \& Begelman 2003; and the analytic models of Blandford
\& Begelman 1999, 2004; Quataert \& Gruzinov 2000).  The accretion
rate onto a black hole thus probably cannot exceed $\sim \dot M_{Edd}$
by a very large factor, even if the mass supply rate is large (see
Shakura \& Sunyaev 1973 for an early discussion of this
point).\footnote{Note that this is fully consistent with the mean
accretion efficiency of $\sim 10$\% inferred from comparing the local
black hole mass density with the integrated quasar light, even though
accretion is not via a thin disk.}

The above discussion focuses on whether highly super-Eddington
accretion is possible.  The question of whether the Eddington limit
can be exceeded by a modest factor of $\sim 10$ is a bit more subtle.
Magnetized radiation dominated accretion disks are subject to a
``photon bubble'' instability that nonlinearly appears to lead to
strong density inhomogeneities (see, in particular, Begelman 2001,
2002; and Arons 1992; Gammie 1998; Blaes \& Socrates 2001).  Density
inhomogeneities allow super-Eddington fluxes from the accretion flow
because radiation leaks out of the low density regions while most of
the matter is contained in high density regions.  Begelman (2002)
estimates that the Eddington limit can potentially be exceeded by a
factor of $\sim 10-100$.  This would allow much more rapid growth of
black holes at high redshifts, circumventing the above arguments that
seed black holes at $z\sim 15$ are required.  Magnetohydrodynamic
(MHD) simulations of radiation dominated accretion flows are in
progress that should help assess the nonlinear saturation of the
photon bubble instability (Neal Turner and collaborators).

\subsection{Accretion versus Mergers}   
\label{zoltansubsec:mergers}

Mergers between halos can help build up the mass of individual black
holes (without significantly changing the total mass of the
population), provided that the central black holes in the halos
coalesce rapidly.  The mean accretion efficiency of $\sim 10$\%
inferred from comparing the local black hole mass density with the
integrated quasar light suggests that accretion dominates at least the
last e-folding of the black hole mass (Yu \& Tremaine 2002).  Mergers
may, however, be significant earlier on (Haiman, Ciotti, \& Ostriker
2004).  In addition, uncertainties in the {\it expected} radiative
efficiency of black hole accretion limit how accurately one can
constrain the growth of black hole mass by mergers. For example, if
the typical efficiency was $\approx 40$\%, as for a maximally rotating
Kerr black hole, then mergers could clearly dominate black hole growth
(on the other hand, note that multiple mergers would have a tendency
to cancel the black hole spin; Hughes \& Blandford 2003).  In order
for mergers to contribute significantly to the growth of individual
black hole masses, stellar seeds must be present in most of the
numerous minihalos that form at $z\gsim 15$, down to small halo
masses.

For concreteness, consider possible merger histories for the $z=5.82$
SDSS quasar SDSS 1044-0125 (Haiman \& Loeb 2001; the following
arguments would be stronger for the more recently discovered $z=6.41$
quasar SDSS J1148+5251).  One can estimate the mass of the dark matter
halo harboring the quasar by its abundance. SDSS searched a comoving
volume of $\sim 1$ Gpc$^3$ to find each quasar. Assuming a duty cycle
of a few times $10^7$~years, one estimates that the dark matter halos
corresponding to this space density have masses of $10^{13}~\msun$
(using the halo mass function in Jenkins et al.\ 2001; the original
Press \& Schechter 1974 formula would give a similar answer).  A
$10^{13}~\msun$ halo at $z=6$ typically has only $\sim 10$ progenitors
with circular velocities of $v>50$~km~s$^{-1}$ (the other progenitors
being smaller).  This implies that mergers can only help build up the
black hole mass if seed black holes are present in progenitor halos
with much smaller masses.  Haiman \& Loeb (2001) argued for a cutoff
in the black hole mass function around halos with $v=50$~km~s$^{-1}$
because the cosmic ultraviolet background can suppress gas infall into
smaller halos (e.g., Efstathiou 1992; Thoul \& Weinberg 1996; Navarro
\& Steinmetz 1997; Kitayama \& Ikeuchi 2000).  However, Dijkstra et
al.\ (2004b) have recently shown that this suppression is ineffective
at redshifts as high as $z~\gsim 6$.  Thus, there is no known obstacle
to forming seed black holes in halos down to $v\sim 10$~km~s$^{-1}$
(below this threshold, atomic H cooling becomes inefficient).  It
therefore needs to be reassessed whether a large fraction of the mass
growth can be accounted for by mergers among halos. Clearly, placing a
seed black hole in each arbitrarily low mass progenitor halo, with the
same black hole mass to halo mass ratio as inferred for the SDSS
quasars ($\mh/M_{\rm halo}\sim 10^{-4}$), could account for the
observed black hole masses in quasars by $z=6$, even without any gas
accretion (Haiman, Ciotti, \& Ostriker 2004).

A promising way of assessing the role of mergers in black hole growth
and evolution is via their gravity wave signatures (see, e.g., Menou
2003 or Haehnelt 2003 for reviews).  In particular, mergers occur
frequently between the dark matter halos that host high redshift black
holes.  If each such merger results in the coalescence of two massive
black holes, the expected event rates by the {\em Laser Interferometer
Space Antenna (LISA)} are significant (see below).

The question of whether halo mergers necessarily lead to black hole
mergers is, however, still not resolved (see Milosavljevic \& Merritt
2004 for a review).  During a galaxy merger, the black holes sink via
dynamical friction to the center of the galaxy and form a black hole
binary on scales of about a parsec.  The black hole binary can
continue to shrink by ejecting low angular momentum stars that pass
close to the binary (those in the ``loss cone'').  In spherical
galaxies, this process is inefficient because the loss cone must be
replenished by two-body relaxation.  The black hole binary thus
appears to stall and cannot coalesce even during a Hubble time (e.g.,
Begelman, Blandford, \& Rees 1980).  Several ideas for circumventing
this difficulty have been proposed.  Gas accretion may drag the binary
together in a manner similar to Type II migration in planetary systems
(Gould \& Rix 2000; Armitage \& Natarajan 2002).  In addition, in
triaxial galaxies, low angular momentum orbits are populated much more
efficiently because the stellar orbits can be chaotic; the resulting
binary decay times are in many cases significantly less than a Hubble
time, even if only a few percent of the stellar mass is on chaotic
orbits (e.g., Yu 2002; Merritt \& Poon 2004).  Finally, if SMBHs are
brought together by successive halo mergers at a rate higher than the
rate at which they can coalesce, then the lowest mass SMBH is likely
to be ejected out of the nucleus of the merger remnant by the
slingshot mechanism (Saslaw, Valtonen, \& Aarseth 1974), with
implications both for gravity wave event rates and for SMBH mass
build-up.

Essentially all of the work on the gravity wave signal from black
hole-black hole in-spiral has assumed efficient (nearly instantaneous)
mergers.  Because {\it LISA} has spectacular sensitivity, it can
detect such mergers at any redshift (if black holes are present). A
more important constraint is on the masses of the merging black
holes---the sum of the two coalescing holes needs to be
$10^3~\msun~\lsim M~\lsim 10^6~\msun$ in order for the resulting
gravity waves to be within {\it LISA}'s frequency range (e.g., Menou
2003).  Several authors (Haehnelt 1994; Menou, Haiman, \& Narayanan
2001; Islam, Taylor, \& Silk 2004; Wyithe \& Loeb 2003a) have made
predictions for {\it LISA} event rates.  If every galaxy hosts a
massive black hole, {\it LISA} should detect several hundred mergers
per year, with most events at high redshifts (Menou, Haiman, \&
Narayanan 2001; see also Wyithe \& Loeb 2003a).  On the other hand,
Kauffmann \& Haehnelt (2000) argue that only galaxies with deep
potential wells ($v_c~\gsim 100$ km/s) will form SMBHs; in this case,
the event rate is much less ($\sim 1$~yr$^{-1}$) and is dominated by
$z~\lsim 5$ (Haehnelt 2003).  On a related note, Menou et al.\ (2001)
showed that the {\it LISA} event rate is very sensitive to the
fraction of dark matter halos that host massive black holes.  This can
be $\ll 1$ at high redshifts (implying a low {\it LISA} event rate)
because mergers ensure that every galaxy will end up with a black hole
at its center by $z=0$, anyway.  Note that in this case, the predicted
redshift distribution of {\it LISA} events would be very different,
peaking at low redshifts ($z\sim 2$ in Fig.~2 of Menou et al.\ 2001).
These examples highlight the fact that gravity waves will provide a
powerful probe of the formation and growth of SMBHs.  We also note
that predictions for the gravity wave ``lightcurve'' have been
published to date only for equal mass black holes (Hughes 1998); since
the typical merger will take place with large mass ratios, it is
necessary to work out predictions for the general case.

Finally, we note that the observed morphologies of quasar hosts can,
in principle, provide constraints on the prevalence of mergers.  In
numerous existing models, quasar activity is exclusively triggered by
mergers; one then expects the images of quasar hosts to appear
disturbed.  Direct interpretation is difficult because galaxies may
relax and display an undisturbed morphology on a timescale shorter
than the lifetime of the activated quasar, especially after minor
mergers (with large mass ratios).  However, it is interesting to note
that hosts appear clumpy at high redshifts, and smoother and relaxed at
lower redshifts (e.g., Kukula et al. 2001), broadly consistent with
the merger rates of dark halos peaking at high redshifts (Haiman \&
Menou 2000; Kauffmann \& Haehnelt 2000).

\subsection{Stars versus Black Holes}
\label{zoltansubsec:collapse}

Instead of growing by accretion/mergers from solar mass progenitors,
SMBHs may form directly in the collapse of gas clouds at high
redshifts, via a supermassive star or disk.  This depends critically on
whether fragmentation of the gas cloud into stars can be avoided,
particularly in view of the large angular momentum barrier that must
be overcome to reach small scales in a galactic nucleus.  

A number of papers have sketched how this may occur (e.g., Haehnelt \&
Rees 1993; Loeb \& Rasio 1994).  The essential idea is that when
contracting gas in a protogalactic nucleus becomes optically thick and
radiation pressure supported, it becomes less susceptible to
fragmentation and star formation.  It is, however, unlikely that
radiation pressure becomes important before angular momentum does,
implying that the gas forms a viscous accretion disk in the galactic
nucleus (fragmentation before the disk forms can be avoided because
the forming fragments would collide and ``coalesce'' before they can
separate into discrete dense clumps; Kashlinsky \& Rees 1983).  On the
other hand, if self-gravitating, the resulting disk is strongly
gravitationally unstable and becomes prone to fragmentation and star
formation (e.g., Shlosman \& Begelman 1989; Goodman 2003).  Whether
this fragmentation can be avoided is unclear. One possibility is to
stabilize the disk by keeping its temperature high, which may be
possible in a virtually metal free high redshift halo.  In particular,
${\rm H_2}$ molecules are fragile and can be easily dissociated by an
early soft ultraviolet background (Haiman, Rees, \& Loeb 1997). If
molecular hydrogen cooling can be suppressed, the gas will lack
coolants and collapse isothermally at a temperature of $\sim 8000$ K
set by atomic line cooling. If molecules are prevented from forming,
the gas may then be unable to fragment into stars and form a $\sim
10^6~\msun$ SMBH instead (this seems difficult to arrange, but options
to achieve this are discussed in Oh \& Haiman 2003).  Bromm \& Loeb
(2003) have carried out numerical simulations of this scenario and
indeed find that if the temperature is kept at $10^4$~K, $\gsim
10^6~\msun$ can condense to scales of $\lsim 1$~pc.  At the end of
their simulations, the gas is still inflowing with no indication of
fragmentation.

Another possibility is that, even in the presence of significant
cooling, angular momentum transport by gravitational instabilities,
spiral waves, bars, etc., can drive a fraction of the gas to yet
smaller scales in the galactic nucleus.  Eisenstein \& Loeb (1995)
argued that this was particularly likely to occur in rare low angular
momentum dark matter halos because the disk could viscously evolve
before star formation commenced.  In addition, even if most of the gas
is initially converted into stars, stellar winds and supernovae will
eject a significant amount of this gas back into the nucleus, some of
which will eventually collapse to smaller scales (Begelman \& Rees
1978).

Although the detailed evolutionary pathways are still not understood,
a possible outcome of the above scenarios is the continued collapse of
some gas to smaller scales in the galactic nucleus.  As the gas flows
in, it becomes optically thick, and the photon diffusion time eventually
exceeds the inflow time.  Radiation pressure dominates for
sufficiently massive objects so that the adiabatic index is $\Gamma
\approx 4/3$.  Radiation pressure may temporarily balance gravity,
forming a supermassive star or disk (SMS; e.g., Hoyle \& Fowler 1963;
Wagoner 1969; see, e.g., Shapiro \& Teukolsky 1983 for a review and
additional references to earlier work).  The SMS will radiate at the
Eddington limit (but see \S~\ref{zoltansubsec:Eddington} above) and continue
contracting.  When the SMS is sufficiently compact ($GM/Rc^2 \approx
10^{-4} M_8^{-1/2}$ for nonrotating stars), general relativistic
corrections to the gravitational potential become important, and the
star becomes dynamically unstable because its effective polytropic
index is $\lsim 4/3$.  For masses $\lsim 10^5~\msun$, thermonuclear
reactions halt the collapse and generate an explosion (e.g., Fuller,
Woosley, \& Weaver 1986), but more massive objects appear to collapse
directly to a SMBH (see Shapiro 2004 for a review; and, e.g., Shibata
\& Shapiro 2002; Saijo et al. 2002 for recent simulations).

\subsection{The Formation of Black Holes in Stellar Clusters}
\label{zoltansubsec:clusters}

The negative heat capacity of self-gravitating stellar systems makes
them vulnerable to gravitational collapse in which the core of the
cluster collapses on a timescale $t_{cc}$ comparable to the two-body
relaxation time of the cluster (Binney \& Tremaine 1987).  If
core collapse proceeds unimpeded, the resulting high stellar densities
can lead naturally to the runaway collisional growth of a single
massive object which may evolve to form a black hole (as in the discussion of
SMSs above).  This process provides an additional route for the direct
formation of SMBHs at high redshifts (or, more likely, intermediate
mass seeds).

Early work suggested that the fate of stellar clusters depends
sensitively on the number of stars in the cluster.  Lee (1987) and
Quinlan \& Shapiro (1990) found that very dense massive stars clusters
($N~\gsim 10^6-10^7$ stars) were required to have successful
core collapse and runaway growth of a single massive object.  In less
massive clusters, core collapse was halted by binary heating, in which
the cluster gains energy at the expense of binaries via three-body
interactions (Heggie 1975; Hut et al. 1992).  Successful core collapse
also requires that $t_{cc}$ is shorter than the timescale for the most
massive stars to evolve off the main sequence (Rasio, Freitag,
G\"urkan 2004; this
requirement implies compact clusters $\lsim 1$~pc in size).  Otherwise,
mass loss from evolved stars and supernovae prevents the core from
collapsing (in much the same way as binary star systems can become
unbound by supernovae).

Recent work has revived earlier ideas that stellar clusters are
subject to a ``mass segregation instability'' that makes even less
massive clusters prone to forming black holes (Spitzer 1969; Vishniac 1978;
Begelman \& Rees 1978).  Because massive stars in a cluster sink by
dynamical friction towards the center (mass segregation), they
invariably dominate the dynamics of the cluster core and can undergo
core collapse on a timescale much shorter than that of the cluster as
a whole (and on a timescale shorter than their main sequence
lifetime).  Portegies Zwart \& McMillan (2002) showed with N-body
simulations that the resulting core collapse likely leads to runaway
merger and formation of a single black hole. G\"urkan, Freitag, \&
Rasio (2004) reached a similar conclusion for much larger $N \sim 10^7$ 
using Monte Carlo simulations.

The above processes provide a promising channel for the formation of
IMBH seeds, which can grow via mergers
and/or accretion to form SMBHs.  For example, Volonteri et al.\ (2003)
and Islam, Taylor, \& Silk (2003) have recently incorporated such early 
black hole seeds into Monte Carlo simulations of the black hole merger 
histories. With reasonable prescriptions for the merging and accretion 
of black holes inside dark halos, these models can account for the 
observed evolution of the quasar luminosity functions at $z<5$ and 
can serve for physically motivated extrapolations to high redshifts 
to describe the first AGN.

It should be noted that IMBHs may have been directly detected using
stellar dynamics in the globular clusters G1 (Gebhardt et al.\ 2002)
and M15 (van der Marel et al.\ 2002; although this object can be
modeled without an IMBH, van der Marel 2004) and/or as ultraluminous
X-ray sources in nearby galaxies (e.g., Colbert \& Mushotzky 1999;
Kaaret et al.\ 2001).  There are, however, viable non-IMBH
interpretations of both the globular cluster (e.g., Baumgardt et al.\
2003) and ultraluminous X-ray source observations (e.g., King et al.\
2001; Begelman 2002).

\section{The Future}
\label{sec:future}

In this section, we briefly summarize the possibility of probing the
continuum and line emission from AGN beyond the current redshift
horizon of $z\sim 6$.  This discussion is necessarily based on models
for how the AGN population evolves at $z>6$.  Such models can be
constructed by assuming that SMBHs populate dark matter halos, e.g.,
in accordance with the locally measured $\mh-\sigma$ relation
(or an extrapolation of the relation to higher redshifts).  While
there is no direct measurement of this relation at high redshifts, 
this assumption is at least plausible. 
There is, e.g., tentative evidence that the relation holds for $z\sim
3$ quasars (this is based on using the H$\beta$/OIII lines as proxies
for black hole mass and $\sigma$, respectively; e.g., Shields et al. 2003), 
and also at $z\sim 6$ (based on the argument in 
\S~\ref{zoltansubsec:sdss}).  No
doubt the observational constraints will improve as both black hole 
masses and velocity dispersions are measured in larger samples of 
distant quasars (e.g., from the SDSS).  Correspondingly, 
extrapolations to high redshifts will be more reliable as the feedback 
processes that regulate black hole growth are better understood.  
Here we summarize predictions from the simplest models.

\subsection{Broadband Detections}

Predictions for the number counts of high redshift AGN have been made 
using simple semi-analytic models for the near-infrared 
(Haiman \& Loeb 1998b) and in the soft X-rays (Haiman \& Loeb 1999b).  
In these early models, the quasar black hole was assumed to have a 
fixed fraction $\sim 10^{-4}$ of the halo mass, shine at the Eddington 
luminosity, and have a duty cycle of bright activity of $t_q\sim 10^6$~years.

In such models, the surface density of sources is very high in the
optical/near-infrared bands, even at $z\sim 10$.  For example, in the
$1-5\mu$m band, the $\sim 1$nJy sensitivity of the {\em James Webb
Space Telescope (JWST)} will allow the detection of an $\sim
10^5~\msun$ black hole at $z=10$ (provided that the black hole shines
at the Eddington limit with the Elvis et al.\ 1994 spectrum).  Surface
densities as high as several sources per square arcminute are
predicted at this threshold from $z~\gsim 5$, with most of these
sources at $z~\gsim 10$ (Haiman \& Loeb 1999a).  We note, however,
that these predictions are very sensitive to the assumed duty cycle of
bright activity.  For example, for $t_q\sim 10^7$~years, or $\mh
\propto M_{\rm halo}^{5/3}$, the $z\sim 10$ counts can be smaller by a
factor of 10-100 (depending on what redshift--dependence is assumed
for the above scaling relation between black hole and halo mass at
high redshift; see Haiman \& Loeb 1998b; Haehnelt, Natarajan, \& Rees
1998; and Wyithe \& Loeb 2003 for related discussion).  It would also
be interesting to detect the host galaxies of ultrahigh redshift AGN,
which should be feasible with {\em JWST}'s sensitivity.  If the
galaxies occupy a fair fraction ($\sim 5\%$) of the virial radius of
their host halos, then a large fraction ($\gsim 50\%$) of them can
potentially be resolved with {\em JWST}'s planned angular resolution
of $\sim 0.06''$ (Haiman \& Loeb 1998a; Barkana \& Loeb 2000).  The
Large Synoptic Survey Telescope (LSST\footnote{www.lsst.org}; Tyson
2002), with a planned capability of going $\sim5$ magnitudes deeper
than SDSS in a $\sim 3$ times larger solid angle, would be an ideal
instrument for studying high redshift quasars in the
optical/near-infrared, provided that it is equipped with a
sufficiently red filter.

In the soft X-rays, the $0.5-2$~keV flux of $2.5\times 10^{-17}~{\rm
ergs~cm^{-2}~s^{-1}}$ reached in a 2~Ms exposure of CDF-North
(Alexander et al. 2003) corresponds to a larger ($\sim 2\times
10^7~\msun$; see Figure 1 in Haiman \& Loeb 1999) black hole at
$z=10$, but nevertheless, thousands of sources are predicted at
$z\gsim 5$ per square degree, and tens per square degree at
$z>10$. This would imply that tens of $z>5$ sources should have been
detectable already in the CDFs, whereas only a handful of potential
candidates, and no confirmed sources, have been found (as discussed in
\S~\ref{zoltansubsec:deep}). In revised models with longer quasar
lifetimes and thus a steeper scaling of $\mh$ with $M_{\rm halo}$,
these numbers can be sharply decreased (Haiman \& Loeb 1998b;
Haehnelt, Natarajan, \& Rees 1998), which can bring the expected
counts into agreement with current non-detections (Wyithe \& Loeb
2003b).

Large numbers of dusty $z>6$ AGN could be detected at mid-infrared
wavelengths ($\sim 10\mu$m). Although we are not aware of predictions
at these wavelengths for AGN, strong dust-enshrouded starbursts that
turn most of the gas into stars would result in large source counts at
longer wavelengths.  Hundreds of galaxies per square arcminute could
be detectable at the $\sim 100$nJy threshold (in the semi-analytic
models of Haiman, Spergel, \& Turner 2003). This flux level can be
reached in an $\sim 10^6$~s exposure with the {\em Spitzer Space
Telescope}.  Depending on actual source counts, confusion may,
however, set a limit of a few $\mu$Jy for the {\it SST}.  The source
confusion limit is difficult to estimate at long wavelengths ($\gsim
10\mu$m), where counts are currently known only to the 100 times
brighter limit of $10^{-5}$ Jy, and confusion calculations are
model--dependent (see, e.g., figure 3 in V\"ais\"anen et al. 2001).
On the other hand, the $\sim 100$nJy flux density is well within the
sensitivity of future high--resolution instruments, such as the {\em
JWST} and the proposed {\em Terrestrial Planet Finder (TPF)}.

The radio sensitivity of the extended Very Large Array and other
forthcoming instruments (e.g., Allen Telescope Array and Square
Kilometer Array) is also promising for detecting AGN beyond $z\sim
6$. Using the updated scaling of black hole mass with halo mass and
redshift from Wyithe \& Loeb (2003b) and assuming the same radio-loud
fraction ($\sim 10\%$) as at lower redshifts, we find that $\sim $ten
$10\mu$Jy sources per square degree should be detectable at $1-10$~GHz
(Haiman, Quataert \& Bower 2004). The identification of these quasars
is a challenge, but should, in principle, be feasible with deep
optical/IR observations.

In addition to direct detection of AGN at very high redshifts, it may
also be possible to detect lower mass seed black holes at comparable
redshifts (or higher).  In particular, a plausible model for gamma-ray
bursts (GRBs) invokes accretion onto a newly formed $\sim 10~\msun$
black hole (the collapsar model; e.g., Woosley 1993).  GRB afterglow
emission may be directly detectable from $z\sim 10-20$ (e.g., Lamb \&
Reichart 2000; Ciardi \& Loeb 2000).  Such afterglows would show up
as, e.g., fading $I$-band dropouts in infrared surveys (which are
under development; Josh Bloom, private communication). Their detection
would open up a new probe of black hole formation and evolution at
high redshifts (as well as a new probe of the IGM along the line of
sight; e.g., Barkana \& Loeb 2004).

In summary, model predictions for the continuum emission of $z>6$ AGN
are very sensitive to how one extrapolates the $\mh-M_{\rm halo}$
relation to $z\gsim 6$.  However, this should be viewed as ``good
news'': (1) large numbers of detectable AGN at these redshifts are
certainly possible, and (2) their detection will put strong
constraints on models for the origin and evolution of the black hole
population.

\subsection{Emission Line Measurements}

The strongest recombination lines of H and He from $5<z<20$ AGN will
fall in the near-infrared bands of {\em JWST} and could be bright
enough to be detectable.  Specific predictions have been made for the
source counts in the H$\alpha$ emission line (Oh 2001) and for the
three strongest HeII lines (Oh, Haiman, \& Rees 2001; Tumlinson,
Giroux, \& Shull 2001).  The key assumption is that most of the
ionizing radiation produced by the miniquasars is processed into such
recombination lines (rather than escaping into the IGM).  Under this
assumption, the lines are detectable for a fiducial $10^5~\msun$
miniquasar at $z=10$.  The Ly$\alpha$ line is more susceptible to
absorption by neutral hydrogen in the IGM near the source but should
be detectable for bright sources that are surrounded by a large enough
HII region so that Ly$\alpha$ photons shift out of resonance before
hitting the neutral IGM (Cen \& Haiman 2000).

The simultaneous detection of H and He lines would be especially
significant.  As already argued above, the hardness of the ionizing
continuum from the first sources of ultraviolet radiation plays a
crucial role in the reionization of the IGM. It would therefore be
very interesting to directly measure the ionizing continuum of any
$z>6$ source.  While this may be feasible at X-ray energies for
exceptionally bright sources, the absorption by neutral gas within the
source and in the intervening IGM will render the ionizing continuum
of high redshift sources inaccessible to direct observation out to
$1\mu$m.  This is a problem if the ionizing sources are black holes
with $M<10^8~\msun$ at $z\sim 10$ (easily detectable at wavelengths
redward of redshifted Ly$\alpha$ in the near-infrared by {\em JWST},
but too faint to see in X-rays).  The comparison of H$\alpha$ and HeII
line strengths can be used to infer the ratio of HeII to HI ionizing
photons, $Q=\dot{N}_{\rm ion}^{\rm HeII}/\dot{N}_{\rm ion}^{\rm HI}$.
A measurement of this ratio would shed light on the nature of the
first luminous sources, and, in particular, it could reveal if the
source has a soft (stellar) or hard (AGN-like) spectrum.  Note that
this technique has already been successfully applied to constrain the
spectra of sources in several nearby extragalactic HII regions
(Garnett et al.\ 1991).

Provided the gas in the high redshift AGN is enriched to solar levels,
several molecular lines may be visible. In fact, CO has already been
detected in the most distant $z=6.41$ quasar (Walter et al.\ 2003).
The detectability of CO for high redshift sources in general has been
considered by Silk \& Spaans (1997) and by Gnedin, Silk, \& Spaans
(2001).  If AGN activity is accompanied by a star formation rate of
$\gsim 30~\msun/$~yr, the CO lines are detectable at all redshifts
$z=5-30$ by the Millimeter Array (the redshift independent sensitivity
is due to the increasing CMB temperature with redshift), while the
Atacama Large Millimeter Array could reveal fainter CO emission.  The
detection of these molecular lines will provide valuable information
on the stellar content and gas kinematics near the AGN.

\section{Conclusions}
\label{sec:conclude}

In this review, we have summarized theoretical ideas and observational
constraints on how massive black holes form at the centers of
galaxies, and how such black holes grow via accretion and mergers to
give rise to the observed population of black holes in the local and
moderate redshift universe.  This remains a poorly understood but
important problem.  In addition to being of intrinsic interest for
understanding the AGN phenomena, sources of gravity waves, etc., there
is strong evidence that the formation and evolution of black holes is
coupled to the formation and evolution of the host galaxy in which the
black hole resides (e.g., the $\mh-\sigma$ relation), and thus to the
cosmological formation of nonlinear dark matter structures (i.e., the
dark halos surrounding these galaxies).  We anticipate that this will
remain a growth area of research in the coming years, with continued
rapid progress on both the observational and theoretical fronts.

\begin{acknowledgments}
The authors wish to thank the editor of this volume, Amy Barger, for
her patience during the preparation of this article.
\end{acknowledgments}

\end{document}